\documentclass[prl,twocolumn,showpacs,superscriptaddress,floatfix]{revtex4}
\usepackage{graphicx}
\usepackage{bm}
\usepackage{dcolumn}
\usepackage[usenames]{color}
\usepackage{ifthen}
\graphicspath{{all_EPS/}}
\newboolean{comments}
\setboolean{comments}{false}
\newcommand{\C}[1]{\ifthenelse{\boolean{comments}}{{{\color{red}{[#1]}}}}{}}
\begin{document}
\newcommand{\Bz}{C$_{6}$H$_{6}$}
\newcommand{\Cp}{C$_{5}$H$_{5}$}
\newcommand{\Cot}{C$_{8}$H$_{8}$}

%
\title{Controlling the Local Spin-Polarization at the Organic-Ferromagnetic Interface}
\author{Nicolae Atodiresei}\email{n.atodiresei@fz-juelich.de}
\affiliation{Institut f\"ur Festk\"orperforschung and  
Institute for Advanced Simulation,
Forschungszentrum J\"ulich and JARA, 52425 J\"ulich, Germany}
\author{Jens Brede}\email{jbrede@physnet.uni-hamburg.de}
\affiliation{Institute of Applied Physics, 
University of Hamburg, 20355 Hamburg, Germany}
\author{Predrag Lazi{\'c}}
\affiliation{Institut f\"ur Festk\"orperforschung and  
Institute for Advanced Simulation,
Forschungszentrum J\"ulich and JARA, 52425 J\"ulich, Germany}
\affiliation{Massachusetts Institute of Technology, 
Cambridge, 02139 Massachusetts, USA.}
\author{Vasile Caciuc}
\affiliation{Institut f\"ur Festk\"orperforschung and  
Institute for Advanced Simulation,
Forschungszentrum J\"ulich and JARA, 52425 J\"ulich, Germany}
\author{Germar Hoffmann}
\affiliation{Institute of Applied Physics, 
University of Hamburg, 20355 Hamburg, Germany}
\author{Roland Wiesendanger}
\affiliation{Institute of Applied Physics, 
University of Hamburg, 20355 Hamburg, Germany}
\author{Stefan Bl\"ugel}
\affiliation{Institut f\"ur Festk\"orperforschung and  
Institute for Advanced Simulation,
Forschungszentrum J\"ulich and JARA, 52425 J\"ulich, Germany}
\date{\today}
\begin{abstract}
By means of \emph{ab initio} calculations and spin-polarized scanning tunneling
microscopy experiments we show how to manipulate the local 
spin-polarization of a ferromagnetic surface by creating a complex energy
dependent magnetic structure. 
We demonstrate this novel effect by adsorbing organic molecules 
containing $\pi(p_z)$-electrons onto a ferromagnetic surface, in which 
the hybridization of the out-of-plane $p_z$ atomic type orbitals with the 
$d$-states of the metal leads to the inversion of the spin-polarization at the 
organic site due to a $p_z-d$ Zener exchange type mechanism. 
As a key result, we demonstrate that it is possible to selectively inject spin-up and
spin-down electrons from the same ferromagnetic surface, an effect which can be exploited in
future spintronic devices.
\end{abstract}
\maketitle
Combining molecular electronics with spintronics represents one of the most exciting 
avenues in building future nanoelectronic 
devices~\cite{Science282_1660,Science307_531,NatMat8_707}.
For example, widely used in spintronic applications, the spin 
valve~\cite{NatMat8_109} is a layered structure of two ferromagnetic electrodes 
separated by a nonmagnetic spacer to decouple the two electrodes and 
allows spin-polarized electrons to travel through it. 
The efficiency of a spin valve depends crucially on the spin injection into and spin
transport throughout the nonmagnetic spacer.
On one side, since organic molecules are made of light elements with weak 
spin-orbit coupling as C and H, their use as spacer materials is very 
promising for transport properties since the spin coherence over time and 
distance is much larger than in the conventional semiconductors present in 
today's devices~\cite{SSC122_181,JApplPhys93_7358,NatMat4_335}. 
On the other side, the spin injection is mostly controlled by the 
ferromagnetic-organic layer interface~\cite{JApplPhys95_4898,NatMat8_115} 
which is responsible for the significant spin loss in devices~\cite{NJPhys11_2009}. 
Therefore, a large effort is made to control the electronic properties 
at the organic-magnetic interfaces and, in this context, the theoretical 
first-principles calculations represent an indispensable tool to understand 
and guide experiments toward more efficient devices.

In this Letter we propose a simple way to manipulate the local spin-polarization
of a ferromagnetic surface by flat adsorbing organic molecules containing 
$\pi(p_z)$-electrons onto it. 
As a consequence, around the Fermi level an inversion of the local 
spin-polarization at the organic site occurs with respect to the ferromagnetic
surface due to a complex energy- and spin-dependent
electronic structure of the organic-metal interface.
The interaction between the molecule and the ferromagnetic surface reveals a mechanism
similar to the $p_z-d$ Zener exchange~\cite{Dielt2002} and enables a selective control 
of electron injection with different spins [i.e. up($\uparrow$) or down($\downarrow$)] 
from the same ferromagnetic surface within a specific energy range near the Fermi level.

\begin{figure}
    {\includegraphics[scale=0.3]{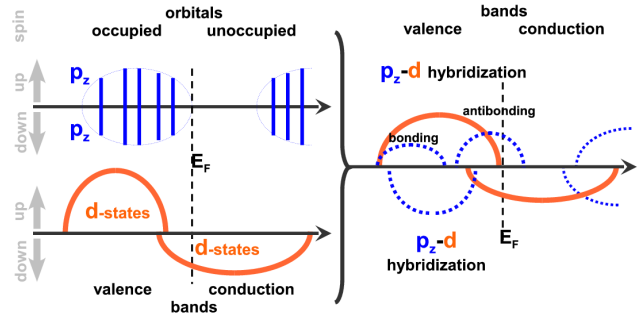}} \\
\caption{(Color online) 
A $p_z-d$ Zener exchange type mechanism explains the interaction between 
nonmagnetic organic molecules with a ferromagnetic surface. 
(left) Cartoons of the molecular electronic structure (upper panel) and 
the density-of-states of the ferromagnetic surface (lower panel). 
(right) The $p_z$ atomic orbitals in the spin-up channel hybridize with the 
majority (spin-up) states of the Fe atoms forming bonding (at lower energies) and 
antibonding (at higher energies) states some of them being pushed above the Fermi 
level. Due to the $p_z-d$ interaction in the spin-up channel, the spin-down 
states are lowered in energy and slightly hybridize with the minority states 
of Fe. As a consequence, for a given energy interval the number of spin-up
states is different than the number of spin-down states.}
\label{fig:1}
\end{figure}
In a first step, we performed {\emph{ab initio}} calculations 
of benzene (Bz), cyclopentadienyl radical (Cp) and cyclooctatetraene (Cot) molecules 
adsorbed on a ferromagnetic 2ML Fe/W(110) surface, a prototypical system used in 
spin-polarized scanning-tunneling microscopy (SP-STM) experiments~\cite{Kubetzka}.
The choice of the calculated molecule-ferromagnetic surface systems can be understood as following:
(i) although having a low reactivity, Bz (C$_6$H$_6$) is an aromatic $6\pi$-electron
system~\cite{Fle78} that can form sandwich type compounds with 
$d$-metals~\cite{BzMBz} and chemisorbs on reactive surfaces~\cite{BzFe100,BzChem}; 
(ii) a high-reactive $5\pi$-electron system, Cp (C$_5$H$_5$) strongly interacts 
with $d$-metals and forms an aromatic $6\pi$-electron system in sandwich type 
molecules like ferrocene~\cite{Fle78}.
However, Cp can be brought on the surface by decomposition of the ferrocene 
molecule which occurs after its adsorption on metallic surfaces~\cite{Cp};
(iii) an $8\pi$-electron system, Cot (C$_8$H$_8$), binds strongly the metal atoms 
by forming an aromatic $10\pi$-electron  system~\cite{Fle78} and is well 
known to react even with $f$-electrons of rare earth 
metals~\cite{Cot-f-Cot,Atodiresei2008} 
forming sandwich type molecules and long nanowires~\cite{Nakajima}.

In a second step, to confirm theoretical predictions as well as
the generality and broad applicability of the $p_z-d$ Zener exchange 
type mechanism, we performed spin-polarized scanning 
tunneling microscopy experiments~\cite{Science309_1542,Science320_82} 
for a phthalocyanine molecule (H$_2$Pc) adsorbed on 2ML Fe/W(110).  
The choice of the H$_2$Pc (C$_{32}$N$_{8}$H$_{18}$), an $40\pi(p_z)$-electron system,
is motivated by its flat structure and the large size 
which makes it an easy observable in experiments~\cite{Science309_1542,Iacovita2008}.

\begin{figure}
    {\includegraphics[scale=0.3]{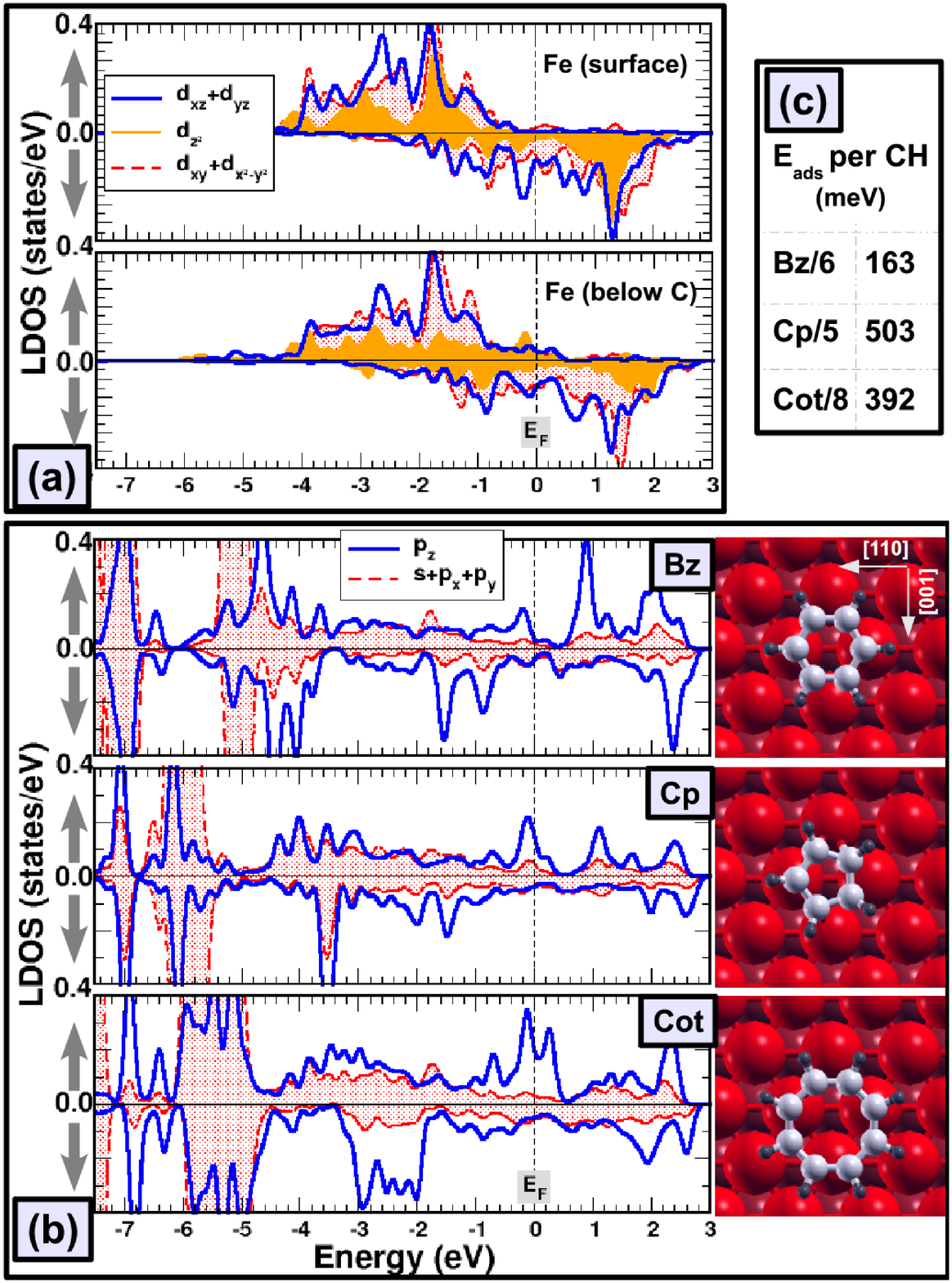}} \\
\caption{(Color online) 
(a) Spin-resolved local density-of-states of an Fe atom of the clean surface 
(upper panel) and an Fe atom below C atom of the Cot molecule (lower panel); 
(b) adsorption geometries and the spin-resolved local density-of-states of the 
Bz, Cp and Cot molecules adsorbed on the 2ML Fe/W(110);
(c) the adsorption energies of the Bz (C$_6$H$_6$), Cp (C$_5$H$_5$) and 
Cot (C$_8$H$_8$) are given in meV per CH group of atoms.
Compared to Bz, Cp and Cot molecules interact strongly with the surface due 
to an effective hybridization between out-of-plane orbitals ($p_z$ of C and 
$d_{z^2}$, $d_{xz}$, $d_{yz}$ of Fe), while the in-plane orbitals are weakly 
interacting ($s$, $p_x$, $p_y$ of C and/or $d_{x^2+y^2}$, $d_{xy}$ of Fe). 
All adsorbed molecules show a general characteristic: the {\emph{energy dependent 
spin-polarization}}, i.e. in a given energy interval the number of 
spin-up and spin-down electrons is {\emph{unbalanced}}. For this 
specific interval the molecule has a net {\emph{magnetic moment}} 
delocalized over the molecular plane, although the total magnetic moment of 
the molecule is $0.0$ $\mu_B$. 
Note also the larger weight of the states crossing the Fermi 
level situated in the spin-down channel at metal site and in the spin-up 
channel at molecular site.}
\label{fig:2}
\end{figure}
Our spin-polarized first-principles calculations are carried out in the 
framework of density functional theory (DFT) by employing the generalized 
gradient approximation (PBE)~\cite{PRL77_3865} in a 
projector augmented plane-wave formulation~\cite{PRB50_17953} 
as implemented in the VASP code~\cite{Kresse1994,Kresse1996}. 
The molecule-Fe/W(110) system is modeled within the supercell approach 
[p(5$\times$3) in-plane surface unit cell] and contains five atomic layers 
(3 W and 2 Fe)  with the adsorbed molecule on one side of the slab~\cite{Makov1995}. 
Using a plane-wave energy cutoff of 500\,eV in our \textit{ab initio} calculations, 
the uppermost two Fe layers and the molecule atoms were allowed to relax 
until the atomic forces are lower than 0.001\,eV/{\AA}.
As a general characteristic of the geometric structure, we note that all molecules 
like to bind with the C and$/$or two C atoms on top of Fe surface atoms so that 
the shortest C$-$Fe bond is about $2.2$ \AA. 
Furthermore, each of the molecules is nonmagnetic~\cite{MAGMOM} upon adsorption 
on the ferromagnetic surface.

\begin{figure}[htb]
    {\includegraphics[scale=0.46]{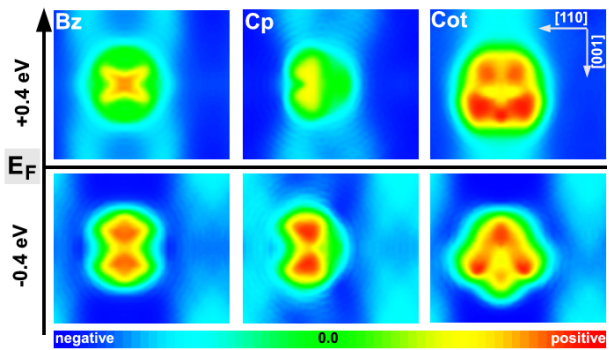}} \\
\caption{(Color online) The spin-polarization at $2.5$ {\AA} 
above the Bz, Cp, and Cot molecules adsorbed on 2ML Fe/W(110) 
($15.85${\AA}$\times13.45${\AA}) surface plotted
for occupied ($[-0.4, 0.0]$~eV) and unoccupied 
($[0.0, +0.4]$~eV) energy intervals around the Fermi level. 
All the organic molecules show a high, locally varying spin-polarization 
ranging from attenuation to inversion with respect to the ferromagnetic Fe film. 
As compared to Bz, the strongly interacting Cp molecule shows an 
amplification of the inversion of the spin-polarization below Fermi level, 
while the Cot molecule shows a strong amplification of the inversion of 
the spin-polarization for both occupied and unoccupied states around the 
Fermi level due to a higher number of $p_z$ interacting electrons.}
\label{fig:3}
\end{figure}
We will first conceptually discuss the mechanism present at the 
molecule-metal interface as schematically depicted in Fig.~\ref{fig:1}. 
The general mechanism, also known as $p_z-d$ Zener exchange mechanism, 
is commonly used to explain how a magnetic impurity in a semiconductor 
host remains magnetic~\cite{Kanamori2001} where, due to the $p-d$ mixing, 
the $p$-band of the semiconductor is broadened and part of it is pushed 
above the Fermi level (often referred as the hole which is created in the 
$p$-band). 
In our specific case, the nonmagnetic molecule represents the impurity 
and its orbitals become spin-split after the interaction with the $d$-states 
of the ferromagnetic surface occured. As shown in Fig.~\ref{fig:1}, in the 
spin-up channel the $p_z$ atomic type orbitals which originally form the 
$\pi$-molecular orbitals hybridize with the majority $d$-states 
of the Fe atoms forming molecule-metal hybrid states with bonding and antibonding 
character. The bonding states are situated at low energies while the antibonding 
states appear at much higher energies, more precisely in an energy window situated 
around the Fermi level. Due to the $p_z-d$ interaction in the spin-up channel, 
the spin-down $p_z$ atomic type orbitals are lowered in energy and slightly hybridize 
with the minority $d$-states of Fe. As a consequence, the states with large weight 
around the Fermi level are in the spin-down channel at clean metal sites and 
in the spin-up channel at the molecule site.
%

The spin-resolved local density-of-states (LDOS) of the molecule-ferromagnetic 
surface systems clearly illustrate the above described mechanism. 
In Fig.~\ref{fig:2} (a) we plot the LDOS of a clean surface Fe atom 
(upper panel) and an Fe below a C atom (lower panel). 
As compared to the clean surface Fe atom, for the Fe below C, the shape of 
the out-of-plane spin-up $d$-states ($d_{z^2}$, $d_{xz}$, $d_{yz}$) is 
strongly changed due to the hybridization with the out-of-plane spin-up 
$p_z$ atomic orbitals of the C.
However, the spin-down $d$-states of iron and $p_z$-orbitals of carbon 
as well as the in-plane states in both spin channels
$d_{x^2+y^2}$, $d_{xy}$ of the metal and $s$, $p_x$, $p_y$ of the molecule
are less affected by the molecule-surface interaction. 
Characteristic for the strongly interacting molecules as Cp and Cot 
[Fig.~\ref{fig:2} (b)] are the spin-up $p_z-d$ bonding states 
in the $[-5.0, -3.0]$~eV energy interval and the spin-up antibondig states 
around the Fermi level in the $[-1, +1]$~eV energy interval. 
In the spin$-$down channel the states having high weight at the molecular site 
are situated at low energies, i.e. $[-3.0, -2.0]$~eV energy interval 
for Cot molecule and around $[-3.5, -1.5]$~eV energy interval for Cp molecule.
Since the Bz-surface interaction is weaker as compared to Cp and Cot molecules 
[see adsorption energies ($E_{ads}$) given in Fig.~\ref{fig:2} (c)], the spin-up 
antibonding molecule-surface hybrid states have smaller weight around the 
Fermi level while the spin-down states at the molecule site are not shifted 
as low in energy (i.e. $[-2.0,-0.8]$~eV energy interval).

The spin-polarized LDOS of the adsorbed Bz, Cp and Cot 
molecules show a very interesting feature: the {\emph{energy dependent 
spin-polarization}}. 
As clearly seen in Fig.~\ref{fig:2} (b) for a given energy interval 
the number of spin-up and spin-down states is {\emph{unbalanced}}. 
For that specific interval the molecule has a net {\emph{magnetic moment}} 
delocalized over the molecular plane since it is carried mostly by 
the $p_z$ states.
This implies that, in a SP-STM experiment for 
which the energy interval is selected by the applied bias voltage, the molecule 
will show a \emph{magnetic contrast} although each of 
the adsorbed molecule is nonmagnetic~\cite{MAGMOM} upon adsorption 
on the ferromagnetic surface. 

Even more interesting, as depicted in the LDOS, around 
the Fermi level the states with high weight are in the spin$-$up channel at 
the molecular site, while on the clean Fe surface these states are in the 
spin-down channel. 
As a consequence, at the molecular site an {\emph{inversion}} of 
the spin-polarization~\cite{SPLDOS} occurs with respect to the 
ferromagnetic surface. 

This effect is clearly illustrated in Fig.~\ref{fig:3} by  the spin-polarization
at $2.5$ {\AA} above the molecule  in the energy intervals 
below $[-0.4, 0.0]$~eV and above $[0.0, +0.4]$~eV the Fermi level. 
A common characteristic for all the molecule-ferromagnetic surface systems 
is the high and locally varying spin-polarization ranging from {\emph{attenuation}} 
to {\emph{inversion}} with respect to the ferromagnetic surface. 
Since the carbon's $p_z$ atomic orbitals of the Cp and Cot molecules are 
strongly interacting with the $d$-states of the iron atoms, as compared to 
Bz molecule, an {\emph{amplification}} of the molecule's local 
spin-polarization below Fermi level occurs. 
Moreover, because of the higher number of $p_z$ electrons present in Cot 
molecule, many antibonding $p_z-d$ hybrid states with high weight above 
the Fermi level are created. Overall, this results in a strong 
{\emph{amplification}} of the molecule's local spin-polarization also above 
the Fermi level (see Fig.~\ref{fig:3}) as compared to Bz and Cp molecules. 

In general, the probability to inject spin-polarized electrons from a 
ferromagnetic electrode is proportional to the density-of-states near the 
Fermi level, i.e. with the spin-polarization in the energy interval defined 
by the applied bias voltage. 
Our first-principles calculations demonstrate that by flat adsorption 
of organic molecules onto the ferromagnetic surface, an {\emph{inversion}} 
of the spin-polarization occurs with respect to the ferromagnetic iron film. 
We conclude that it is possible to control locally the injection of 
electrons with different spins (i.e. up or down) from the same ferromagnetic 
surface by flat adsorbing organic molecules containing $\pi(p_z)$-electrons onto it. 
With other words, from the clean ferromagnetic surface mostly the  spin-down 
($\downarrow$) electrons will be injected while locally, at the molecule site, 
mostly the spin-up ($\uparrow$) electrons are injected.
\begin{figure}[htb]
    {\includegraphics[scale=0.38]{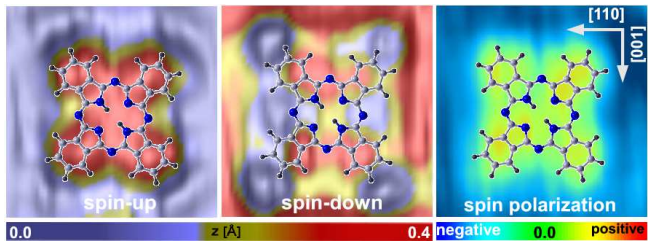}} \\
\caption{(Color online) 
Experimental ($22$\AA$\times22$\AA) SP-STM images for H$_2$Pc 
adsorbed on 2ML Fe/(W110)
at $U = +0.05$~V for both spin channels
[i.e. up ($\uparrow$) and down ($\downarrow$)] and
local spin-polarization.
H$_2$Pc molecules show a high, locally
varying spin-polarization ranging from attenuation to inversion
with respect to the ferromagnetic Fe film.
}
\label{fig:4}
\end{figure}

To validate the predictions of the {\emph{ab initio}} calculations and to 
demonstrate that the mechanism explaining the organic-ferromagnetic metal interface 
discussed in Fig.~\ref{fig:1} is {\emph{general}} and {\emph{widely}} applicable
for other flat organic molecules which contain 
$\pi(p_z)$-electrons, we have performed SP-STM experiments~\cite{SuppInfo}
on the large-size phthalocyanine molecule (H$_2$Pc) adsorbed on 2ML Fe/(W110). 
Due to its $\pi(p_z)$-electrons, the H$_2$Pc molecule has a flat structure and 
is well known for its high reactivity similar to the one of the Cot molecule,
i.e. H$_2$Pc reacts with 
(a) $d$-metals forming metal-phthalocyanine (MPc) molecules~\cite{Fle78} or 
(b) $f$-metals yielding sandwich type organometallic compounds~\cite{Pc-f-Pc}. 

The experimental SP-STM images recorded for the H$_2$Pc adsorbed on 
2ML Fe/(W110) (see Fig.~\ref{fig:4}) clearly show that the H$_2$Pc 
molecule exhibits a magnetic contrast in the $[0.0 , 0.05]$~eV 
energy interval above Fermi level (i.e. the molecular 
appearance of the spin-up and spin-down channels is obviously different).
As a consequence, in this specific energy interval, the H$_2$Pc molecule
has a net magnetic moment delocalized over the molecular plane as depicted 
in Fig.~\ref{fig:4} by the experimentally determined local spin-polarization~\cite{PRL65_247}.

To summarize, we have shown that the spin-polarization of a ferromagnetic surface
can be locally {\emph{inverted}} by flat adsorbing organic molecules 
containing $\pi(p_z)$-electrons onto it. 
The complex energy dependent magnetic structure created at the organic 
molecule-surface interface resembles the $p_z-d$ exchange type mechanism. 
Although the adsorbed molecules are nonmagnetic, due to an energy dependent 
spin-polarization, in a given energy interval the molecules have 
a net magnetic moment delocalized over the molecular plane.
Highly electronegative molecules as Cp and Cot 
strongly interact with the ferromagnetic surface which yields to an
amplification of the inversion of the spin-polarization as compared to less reactive molecules as Bz. 
Moreover, the generality of the theoretical results and the $p_z-d$ exchange 
type mechanism is further demonstrated by our SP-STM experiments on the 
H$_2$Pc adsorbed on the 2ML Fe/(W110) surface which clearly show the inversion
of the local spin-polarization near the Fermi level. 
Our combined first-principles and experimental study demonstrates that electrons 
of different spin [i.e. up ($\uparrow$) and down ($\downarrow$)] can selectively be 
injected from the same ferromagnetic surface by locally controlling the inversion 
of the spin-polarization close to the Fermi level, an effect which can be exploited
to increase the efficiency of future molecular spintronic devices.

{\small{This work is funded by the DFG (SPP1243, SFB 668-A5 and GrK 611). 
The computations were performed on JUROPA and JUGENE supercomputers at the 
 J\"ulich Supercomputing Centre, Forschungszentrum J\"ulich (Germany).}}

\begin{thebibliography}{100}
\bibitem[{\citenamefont{SPLDOS}(0000)}]{SPLDOS}
\bibinfo{author}{\bibfnamefont{
The spin-polarization is defined as
$(n^{\uparrow} - n^{\downarrow}) / (n^{\uparrow} + n^{\downarrow})$, 
where $n^{\uparrow}$ and $n^{\downarrow}$ are the spin-up and spin-down
charge densities in a given energy interval
}}.
\end{thebibliography}
%

\end{document}